\documentclass[10pt,sigconf,letterpaper]{acmart}

\renewcommand\footnotetextcopyrightpermission[1]{} 
\usepackage{booktabs} 

\usepackage[utf8]{inputenc}

\usepackage{color,hyperref,graphicx}
\usepackage[T1]{fontenc}
\usepackage{textcomp}

\newcommand{\todo}[1]{{\color{red} #1}}
\newcommand{\CD}[1]{{\color{blue} Clayton: #1}}

\newcommand{\proposeddelete}[1]{{\color{blue} Clayton proposes deleting this block: #1}}
\renewcommand{\proposeddelete}[1]{} 

\begin{document}

\title{Investigating the effectiveness of web adblockers}

\author{Clayton Drazner}
\authornote{Now working at Google}
\orcid{1234-5678-9012}
\affiliation{%
  \institution{Rice University}
}
\email{drazcmd@gmail.com}

\author{Nikola Đuza}
\affiliation{
  \institution{Faculty of Technical Sciences, University of Novi Sad, Serbia}
}
\email{nikolaseap@gmail.com}

\author{Hugo Jonker}
\affiliation{%
  \institution{Open Universiteit, Heerlen, Netherlands}
}
\email{hugo.jonker@ou.nl}

\author{Dan S. Wallach}
\affiliation{%
  \institution{Rice University}
}
\email{dwallach@rice.edu}


\begin{abstract}
We investigate adblocking filters and the extent to which websites
and advertisers react when their content is impacted by these filters.
We collected data daily from the Alexa Top-5000 web sites for 120 days, and from
specific sites that newly appeared in filter lists for 140 days. By evaluating
how long a filter rule triggers on a website, we can gauge how long it
remains effective. We matched websites with both a regular adblocking
filter list (\emph{EasyList}) and with a specialized filter list that
targets anti-adblocking logic (\emph{Nano Defender}).
From our data, we observe that the effectiveness of the EasyList
adblocking filter decays a modest 0.13\% per day, and after
around 80 days seems to stabilize. We found no evidence
for any significant decay in effectiveness of the more specialized, but
less widely used, anti-adblocking removal filters.
\end{abstract}


\maketitle

\section*{KEYWORDS}
Adblocking, anti-adblocking, web crawling.
%


\section{Introduction}

Internet advertising is a significant source of income for many web
sites as well as for apps on mobile platforms. On the other hand,
advertisements are not desired by many users. Ads consume bandwidth and
battery power, and are a source of attacks, such as
malvertisements and social engineering hacks (e.g., ``viruses found on your
system!''). Adblockers help users block this unwanted content. Early
adblockers were simple DNS blacklists. Modern adblockers follow a
standardized syntax to specify patterns in hostnames as well as digging
deeper into the DOM structure of a website to remove specific elements (see Section~\ref{sec:how-adblocking-works}).
A standardized syntax allows anyone to create filter rules. These rules
can be generic to any web site while others are
site-specific or even country-specific. EasyList and other organizations
centrally publish lists of these rules, which can be imported by
a variety of different ad-blocking extensions supported by most modern web browsers.

Such lists threaten advertising and the accompanying revenue.
Unsurprisingly, both advertisers and their hosting websites have found a
variety of ways to push back. Recent anti-adblocking technologies allow
websites to detect the presence of adblocking systems and change
their behavior accordingly. Websites can then request the user to
disable their adblocker, or simply block the detected adblock-using
visitor. This has the makings of an ``adblock war'': users are blocking
ads, and websites are blocking users. A new advertisement may initially
bypass the filters, which will cause the filter lists to be
updated and block the ad\ldots which may cause the advertiser to update
its code to defeat the blocking.

In this paper, we investigate whether filter lists and advertisers
respond to each other in such a fashion. For three months, we
continuously monitored filter lists and the websites affected by them,
adding new websites as filter lists expanded. For each rule in a list,
we tracked when it affected the corresponding website. We looked at the
aggregated data to see if there was an overall trend, that is, how long
it took for advertisers to react to a new blocking rule, and how long it
took for filter rules to react to updated advertisements.

We will look at two separate filter lists:
EasyList\footnote{\url{https://EasyList.to/}}, which is widely used and
enabled by default in many browser ad-blocking plugins, and
Nano Defender\footnote{%
\url{https://jspenguin2017.github.io/uBlockProtector/}}, which is much
less popular but specifically targets web page logic that attempts to
defeat ad-blocking.

\ \\
\textbf{Contributions and limitations:}
When we began this work, no other study had used daily scans of large
numbers of websites to analyze adblocking. Now, several other researchers
have conducted studies in this area (discussed in Section~\ref{sec:related}),
but with different methodologies.

Our study has several limitations. First, we use filter lists respecting
the standard ``Adblock Plus'' syntax. We do not measure other
adblocking or privacy-enhancing techniques, such as
behaviors of or responses to browser extensions such as Ghostery or
NoScript. Moreover, our study's results are confined to the capabilities
of the Adblock Plus engine and the coverage of the monitored lists: any
advertisement not blockable by the engine, or not in any of the lists,
is not considered. Lastly, there are several possible sources of noise
in our data. Some of these are internal and known, such as failure to
connect to a website on a specific day. There may also be external
factors, such as a website switching ad providers.

\section{Background Information}
\subsection{Related work}
\label{sec:related}
Many studies have analyzed online advertising and adblocking. Some
examined overall adblock usage and/or sentiment, either by surveying
users~\cite{CC15,MVNC18} or analyzing aggregate user
data~\cite{MMCB16,PHF15}. Others have looked at the effectiveness of
adblocking as a privacy tool~(e.g., \cite{BW2018,FBLS18,TTGMM17,WU16}).
A few have also examined topics such as browsing ``quality of
experience''~\cite{NB19} and performance~\cite{GKM17}. Several studies
took a more direct approach to detecting ads, such as automatically
sourcing adblocker filter rules~\cite{GHAL15}, detecting ad-blockers
through inspecting network traffic~\cite{MBMC18}, foiling ad-blocker
detection~\cite{bruguera17}, and perceptually detecting
ads~\cite{TDRPB18,TDRPB19}).

The overarching adblock vs.~advertiser arms race has also been studied.
Storey et al.~\cite{SRMN17} characterize three stages in the arms race:
first, adblocking by users versus ad obfuscation by advertisers; second,
adblock detection by advertisers versus obfuscated adblocking by users;
and last, blocking adblock detectors by users versus obfuscated adblock
detectors by advertisers. Gritckevich et al.~\cite{GKS18} examine
adblockers from a game theory perspective, developing a model to examine
how adblockers affect users and ad publishers. Mughees et
al.~\cite{MQSDH16,MQS17} use automated A/B testing to determine the
incidence of anti-adblocking techniques on the web. They scanned the
Alexa Top 100K and found at least 0.7\% of sites use adblock-detection
techniques, asking their users to turn off adblocking once detected.
Nithyanand et al.~\cite{NKJ16} study the prevalence of adblock detection
techniques by focusing on third-party services in the Alexa Top 5K
websites, finding that at least 6.7\% of the sites in their sample used
adblock detection techniques. Zhu et al.~\cite{ZHQ18} created an
``anti-adblock'' detection approach. They automatically visit targeted
sites multiple times with and without adblocking and analyze the
differences in the execution time. They found that 30.5\% of the
Alexa Top 5K used some form of anti-adblocking code.

Parts of the work by Iqbal et al.~\cite{ISQ17} resemble our
study. In 2017 they set out to determine how adblock detection evolved.
They approach this by using the Internet Archive to retrieve previous
versions of websites and match them against a filter list. Our study
uses similar adblock detection, but we collected live data directly, not
archived data via a third party. In 2018, they published a followup
study~\cite{ISSZQL18} looking at multiple layers of the web stack (HTML,
HTTP, Javascript), building a supervised machine learning model to block
ads and trackers.

A recent tech report by Vastel et al.~\cite{VSL18} independently uses a
similar approach to ours.
They also analyzed EasyList's performance on Alexa top sites in order to
better understand filter lists. While doing so, they posed several of
the same questions we do in this paper and used a number of methods that
were similar to our own, albeit relying on a substantially different
implementation approach. Our work on this area differs in two key ways
from the work by Vastel et al. First, we examined more filter lists
than just EasyList. Second, we use different statistical methods
to summarize our data.

\subsection{How adblocking works}
\label{sec:how-adblocking-works}

Adblocking filters have rules expressed in a simple
syntax\footnote{\url{https://adblockplus.org/filters}}.
Rules can simply specify a URL or domain name, with or without
wildcards, which will prevent undesired elements from even being loaded.
Rules can also specify DOM elements by ID or class, or by any of a
variety of features, including a path through the DOM tree, or styling
attributes like width and height. Such advertising elements will be
removed from the DOM even if they're added dynamically by JavaScript
behaviors. Because these rules have the potential to be too broad,
exception rules are supported, where an exception will override
a blocking rule.

Anti-adblocking techniques generally work by introducing ``bait''
elements into the DOM that would be removed by an adblocker. By using
JavaScript to inspect the DOM, any missing bait elements imply the
presence of an adblocker. The site can then take additional actions,
such as requesting the user to disable their adblocker. Bait elements
can be avoided using exception rules. Alternatively,
anti-anti-adblocking rules can directly target the JavaScript used by
the anti-adblockers. EasyList has a specific policy with regard to
anti-adblocking rules: ``Anti-Adblock should only be challenged if
the system limits website functionality or causes significant disruption
to browsing''\footnote{\url{https://easylist.to/pages/policy.html}},
whereas Nano Defender has no such limitations.


\section{Methodology}
\label{sec:methodology}

The goal of our experiment is to determine if and to what extent
websites respond to adblock updates and adblockers respond to website
updates. A meaningful answer to this question must be based on data
collected from many websites and filter rulesets taken over time.
Therefore, our methodology is described in two parts: data collection and
data analysis.

\subsection{Data collection}
\label{sec:data-collection}

We collected two main types of data: daily iterations of the EasyList
and Nano Defender (and predecessors) filter lists and daily scrapes of
various targeted websites. For the filter lists, all the relevant files
and commit histories are available on either GitHub or the AdblockPlus
team's Mercurial repository\footnote{\url{https://hg.adblockplus.org}}.
We iterated through every commit in the project histories for every
particular filter list and then downloaded the final revision on every
day there was at least one commit.

We note that Nano Defender effectively is a fork of an earlier project 
called Anti-Adblock Killer, apparently abandoned by its author in 2016.
The Anti-Adblock Killer rules, inherited by Nano Defender, appear
as a single large commit in the Nano Defender ruleset. As these rules are all
significantly older than the time period of our web scraping activities
we ignored them when looking specifically at the day-by-day impact of
a given rule on our corpus of web scrapes.

Our approach to scraping websites differs from most studies discussed in
Section~\ref{sec:related}. Generally speaking, most studies used
Selenium to drive browsers running on local machines. Although doing so
is highly effective for doing one-off jobs, we were performing daily
scrapes on a large number of websites for months on end. We used the
Scrapinghub cloud
platform\footnote{\url{https://scrapinghub.com/platform}}. Using their
open-source Scrapy (configurable web scraper) and Splash (headless
browser) libraries, we were able to easily scrape thousands of websites
per day using one Scrapy cloud unit and a small Splash instance (which
respectively cost \$9 and \$25 USD per month).

At every website we visited, we first scrolled to the bottom and waited
1.5~seconds, to allow for delayed behaviors. We saved a copy of the
page's DOM at this point as well as its HTTP Archive
(HAR)\footnote{\url{http://www.softwareishard.com/blog/har-12-spec/}}
for subsequent processing. During an approximately 140~day timespan, we
collected roughly 487~GB of total data: close to 400~GB from the Alexa
websites (collected during 120 days) and the remainder from websites
specifically targeted by filter rules in Nano Defender list (collected
during 140 days).

We scraped the Alexa Top 5K
on a daily basis for
120 days, as well as websites targeted by the Anti-Adblock Killer
and Nano Defender filter lists for 140 days. We used a Rake
task\footnote{\url{https://github.com/ruby/rake}} that downloads any
commits to the two filter lists on a daily basis and extracts the URLs
from newly added rules. This ensures a short turnaround time between
addition of a new filter rule for a new website, and that site's
inclusion in our scraper.

Our entire data collection process ran roughly from April through August in 2017
with a variety of false-starts and engineering challenges beforehand to get it running.

\subsection{Data processing and analysis}
For each web site image scraped, and for each set of filter rules against which we need to evaluate it, 
we used the open source Libadblockplus
library\footnote{\url{https://github.com/adblockplus/libadblockplus}} to
determine whether that day's filter list would ``trigger''
on the downloaded version of a website for that day. We specifically
chose Libadblockplus because it is a C++ wrapper around the Javascript
Adblock Plus core engine, reducing the likelihood of its behavior differing
from in-browser adblocking. Libadblockplus, when given the result of the web scraper and
a filter list, returns a list of matched web page elements with their corresponding matched rules. An exception rule 
that also matches the same web page element will suppress that web page element from the list of results.

Recall that EasyList and Nano Defender use exception rules in different ways. EasyList uses exceptions
to narrow otherwise overbroad positive rules, avoiding undesired damage to a web page. Nano Defender, on
the other hand, uses exceptions to avoid touching bait elements used by anti-adblocking logic. As such,
we invert the sense of Nano Defender's exception rules; if a Nano Defender exception rule triggers on
a web page element, we consider that to be a successful match, because it's operating to defeat an anti-adblocker.

Due to the size of our data set, we split
these jobs into smaller chunks (generally 250 websites at a time) which
we ran in parallel on our institutional cluster. This enabled us to
process months of scrape data using canonical versions of the various
filter lists in a handful of days. We saved the results of these compute
jobs as simple JSON files, in which we mapped websites to lists of
(filter rule, offending resource) pairs for each day of scraped data.

\begin{figure} 
\includegraphics[width=.5\textwidth]{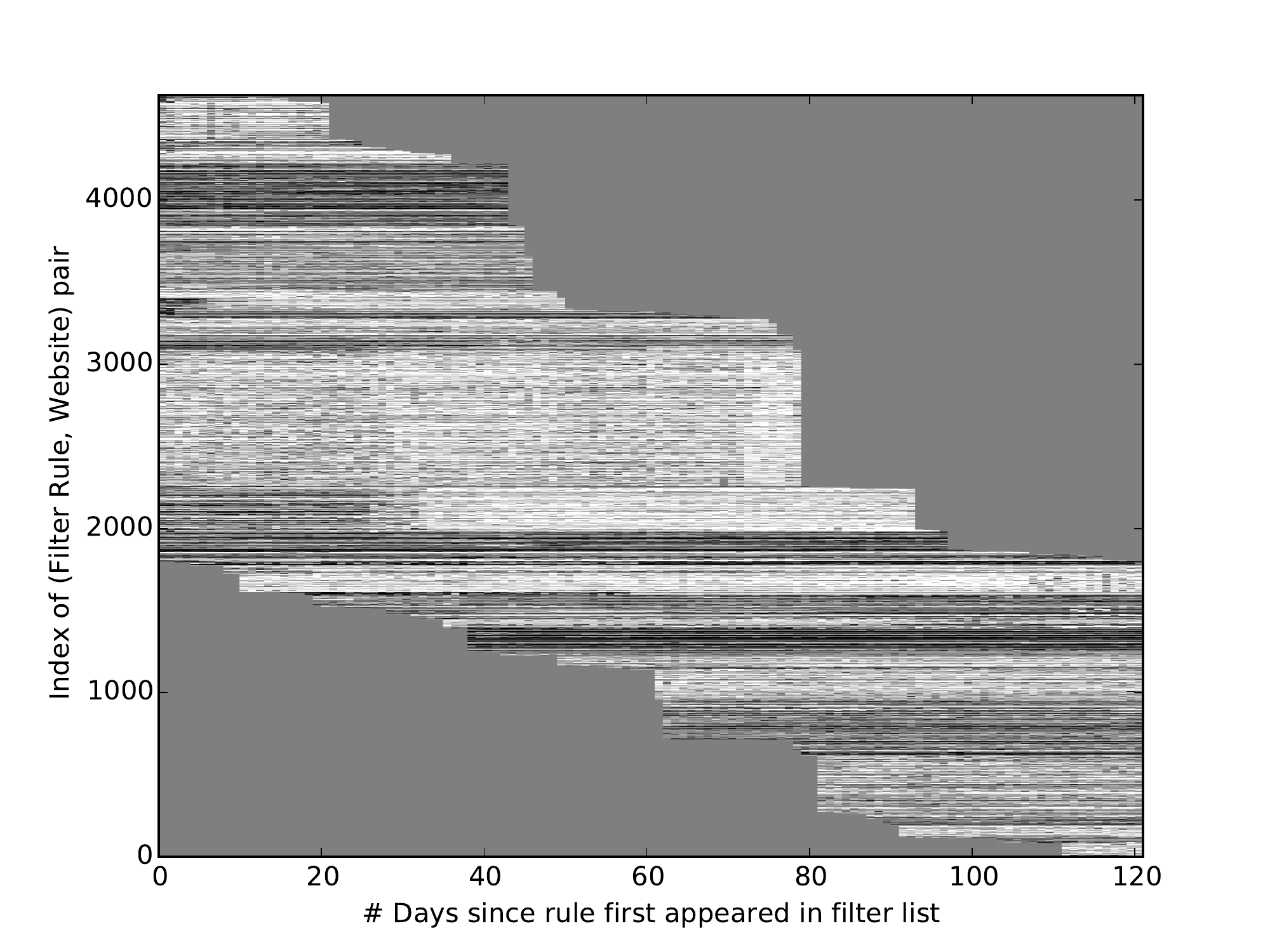}
\caption{EasyList effectiveness on Alexa Top 5K (raw data, rasterized)}
\label{fig:EasyListVsAlexaNoExceptions:raw}

\includegraphics[width=.5\textwidth]{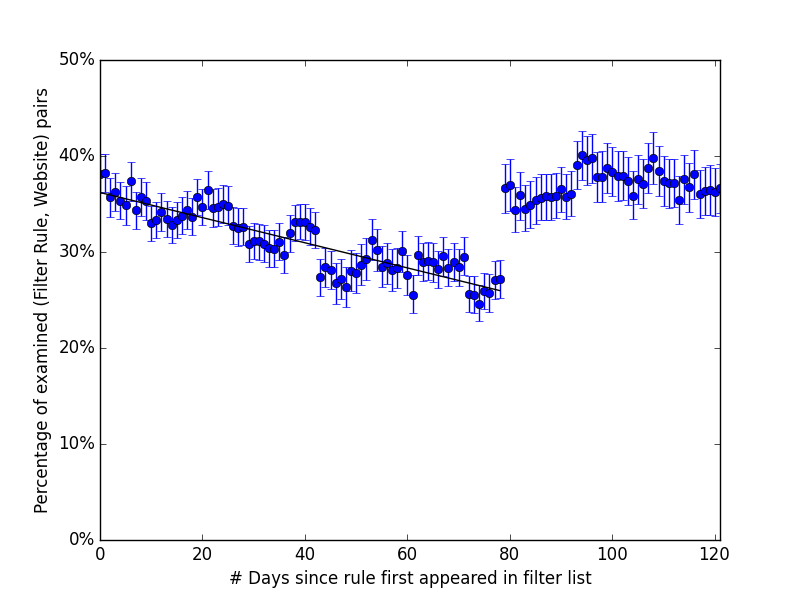}
\caption{EasyList effectiveness on Alexa Top 5K (average with confidence
	intervals)}
    \label{fig:EasyListVsAlexaNoExceptions:avg}
\end{figure}

We plotted these results (see e.g.,
Figure~\ref{fig:EasyListVsAlexaNoExceptions:raw}) with the time since the rule's
introduction on the horizontal axis and collected data on the vertical
axis. Specifically, each row corresponds to one (website, filter-rule)
tuple. For a given day $x$ and a given tuple $y$ (website,
filter-rule), the value of the point $(x,y)$ in the graph is either
{\em true}, {\em false}, or {\em fail}, and colored as follows:

\begin{itemize}
\item Black: {\em true}---the archived copy for day $x$ of the website
	triggered a hit on the filter rule.
\item White: {\em false}---the archived copy for day $x$ did \emph{not}
	trigger a hit.
\item Gray: {\em fail}---no data available.
\end{itemize}

In our graphs and data processing, rows with filter rules that never
triggered on those sites are omitted. As such, for each combination of a
website and a filter rule that triggered at least once on that website,
we ended up with a row of data in our graphs. Failures may be due to
failure to contact the website, but also occur when the day is outside the
observation window of the website. We discuss this further, below.

Our graphing technique is designed to align each row based on the date
at which the rule was introduced.
Each value of $(x, (y_{\text{site}}, y_{\text{rule}}))$ shows
whether filter rule $y_\text{rule}$ still triggered on the copy of website
$y_\text{site}$ after $x$ days.

To clarify this alignment process, consider Figure~\ref{fig:EasyListVsAlexaNoExceptions:raw},
where there are two gray ``triangles'' of missing data. The data rows adjacent to the 
lower-left triangle correspond to cases where the rule {\em predates} the start of our experiment. So if a rule 
was 80 days old at the start of the experiment, then we would only render results for
$x=80 \rightarrow 120$. The data rows adjacent to the upper-right triangle
correspond to cases where the rule appeared while our experiment was ongoing. So, if a rule
appeared on day 80, we would only have 40 days of results for the effectiveness of that
rule, appearing at $x=0 \rightarrow 40$.

A consequence of this alignment process is that {\em vertical} slices through the graph contain
all the filtering effectiveness data we have for rules {\em of a given age}.

\subsection{Statistical Methods}
\label{sec:error-bars}
Plotting the data as described above allows us to
gauge the effectiveness of filter list rules over time. Specifically,
we use the ratio of total number of hits for a given day $x$
versus the total number of hits $+$ misses as an approximation of the
overall effectiveness of filter list rules after $x$ days. We would
expect this ratio to decline as websites take measures to reduce the
effectiveness of filter rules.

Note that collapsing an entire column into a single value misses out on
sources of uncertainty hidden in the value. The process of scraping
websites includes a number of sources of noise. Principal amongst these
is that the number of filter rules for which we do not have data varies
from day to day. For example, in
Figure~\ref{fig:combinedAbpUbpVsAlexa:raw}, there is little data
available after 40 days, and scant after 80. Thus, it is essential to
model the uncertainty in our data.
\proposeddelete{
Though the matching is either true or false,
modern websites on the Internet include large amounts of varying,
non-static content.
This results from real time ad-bidding networks, as
well as dynamic mechanisms that load page content or ads only upon
scrolling. Some web sites even deliberately conduct live A/B testing on
their users.}
\proposeddelete{\todo{possibly insert a sentence about website
changing infrastructure/ad providers etc}}




To get meaningful error bars for our graphs, we cannot simply take the standard deviation of
a sequence of 1's and 0's. Instead, we use \emph{bootstrap
resampling}
(see, e.g., \cite{BMSECI}), a standard technique for
computing many common statistics. In our case, we compute the 95\%
confidence interval over the mean. Bootstrap resampling produces a
robust result without requiring the data to be
normally distributed~\cite{BFPC15}.

\textbf{Bootstrap resampling in a nutshell:} Bootstrap resampling
is a statistical measure that relies on sampling with replacement.
Consider a data set with $n$ elements over which the average is
computed. To determine the confidence intervals, we first randomly
sample (with replacement) $n$ values from the data set. We compute the
average of this sample. We then repeat this sampling process 5,000 times,
yielding 5,000 so-called ``resampled averages''. These are then sorted
from small to large. The 2.5th percentile and the 97.5th
percentile of this list then provide us with bounds on the confidence
interval. This corresponds to a 95\% confidence interval around the
average. We then can render these confidence intervals as error bars on our graphs.

\section{Analysis}

We first consider the popular EasyList filter and next look at the more
targeted Nano Defender filters. We also consider how our data can
confirm results from prior studies.

\subsection{EasyList}
Our analysis of the effectiveness of EasyList filter rules showed two
discontinuous sections (see
Figure~\ref{fig:EasyListVsAlexaNoExceptions:avg}). This is due to the
denominator changing significantly after $x=75$, which is an
artifact of our methodology. As can be seen in
Figure~\ref{fig:EasyListVsAlexaNoExceptions:raw}, we have significantly
more data for the first 75 days of filter rules existence than for the
latter 45 days. This is because EasyList added a large number of filter
rules for new websites on day 45 of our experiment. Since we could
monitor the performance of these rules from their introduction, these
are all plotted from $x=0$ in
Figure~\ref{fig:EasyListVsAlexaNoExceptions:raw}.

In Figure~\ref{fig:EasyListVsAlexaNoExceptions:avg}, the first section
depicts a long period in which large number of
EasyList's filter rules show declining effectiveness. A simple linear
best fit of this section (plotted as a straight line on the graph) shows
a decrease of just over 0.13\% per day, with a total loss of roughly
10\% in about 75 days. These results exclude EasyList's ``exception
rules'' (see Section~\ref{sec:how-adblocking-works}). Taking these rules
into account, we see a the decrease of roughly 0.2\% per day on average,
with a total loss of roughly 15\% in the same period of time. (Visually,
the resulting graphs appear similar to
Figure~\ref{fig:EasyListVsAlexaNoExceptions:avg} and are omitted for
space.) The obvious explanation is that websites and advertisers do
indeed respond to the introduction of new EasyList filter rules.



Following the discontinuity at approximately day 75 we
observe no significant decline; attempting to plot a linear best fit as
we did in the prior section of the graph results in a horizontal line. This
suggests the underlying process of website operators and advertisers
responding to EasyList is not a linear process. A longer time period of
data collection over many more websites might well show a fit to an
exponential curve.


\begin{figure}
\includegraphics[width=.5\textwidth]{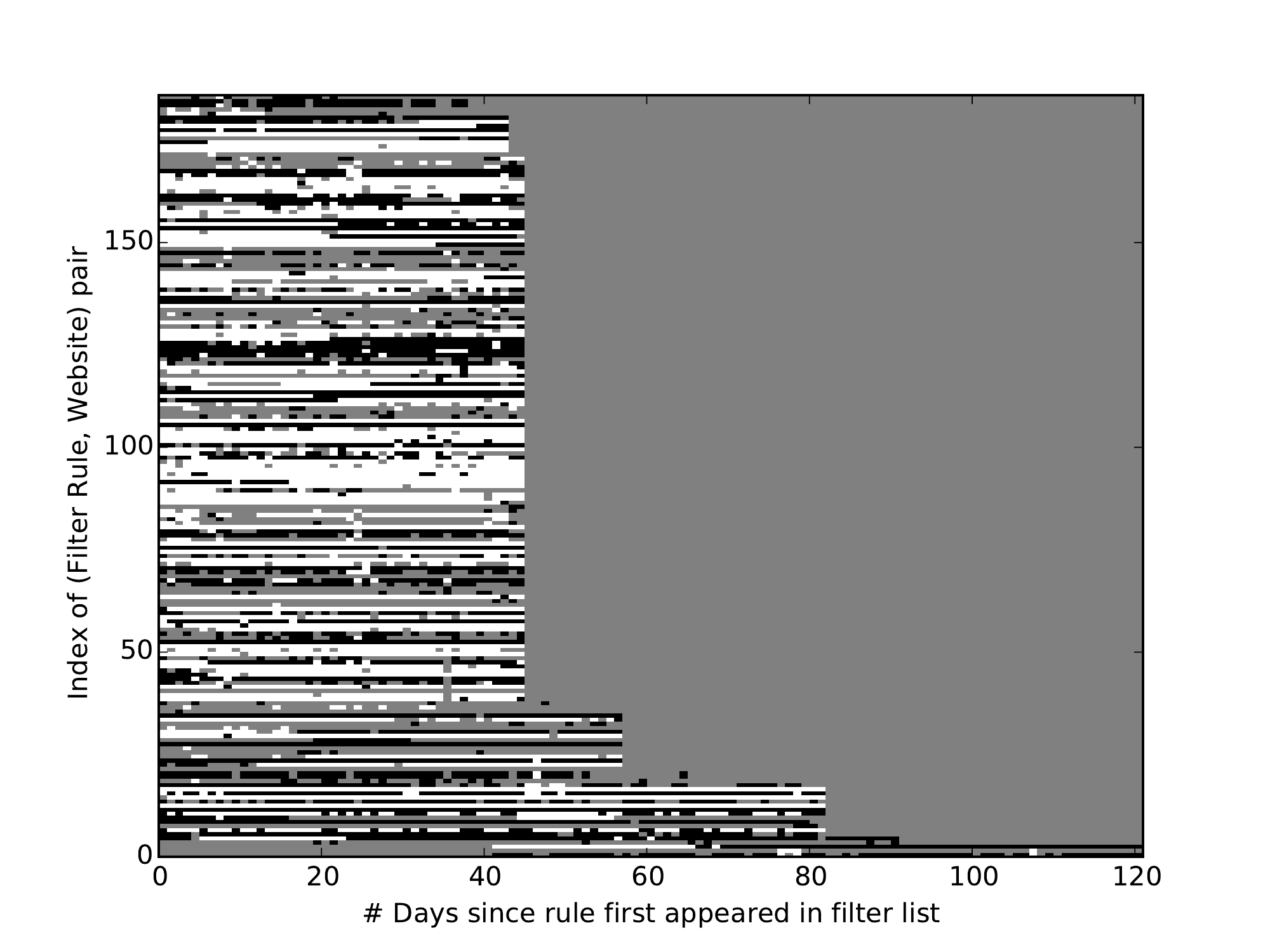}
\caption{Nano Defender effectiveness on the Alexa Top 5K (raw data)}
\label{fig:combinedAbpUbpVsAlexa:raw}

\includegraphics[width=.5\textwidth]{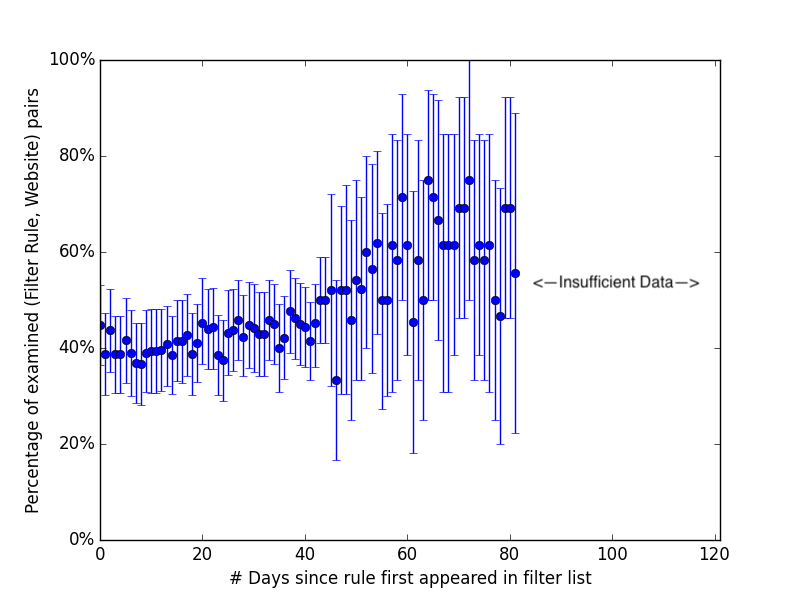}
\caption{Nano Defender effectiveness on the Alexa Top 5K (average with
confidence intervals)}
\label{fig:combinedAbpUbpVsAlexa:avg}
\end{figure}

\proposeddelete{
\CD{ACTUALLY I THINK THIS PART BELOW IS FLAT OUT WRONG. There's
'thin' data, but not a discontinuous section similar to fig. 1. So
if anything, would say something about how the data thins out after
a certain amount of time. But I think that's pretty obvious anyways
looking at the graph itself.}
These graphs, \ref{fig:combinedAbpUbpVsAlexa:raw} and
\ref{fig:combinedAbpUbpVsAlexa:avg}, show discontinuous sections similar to
those in Figure~\ref{fig:EasyListVsAlexaNoExceptions:raw}.}

\begin{figure}[htbp]
\includegraphics[width=.5\textwidth]{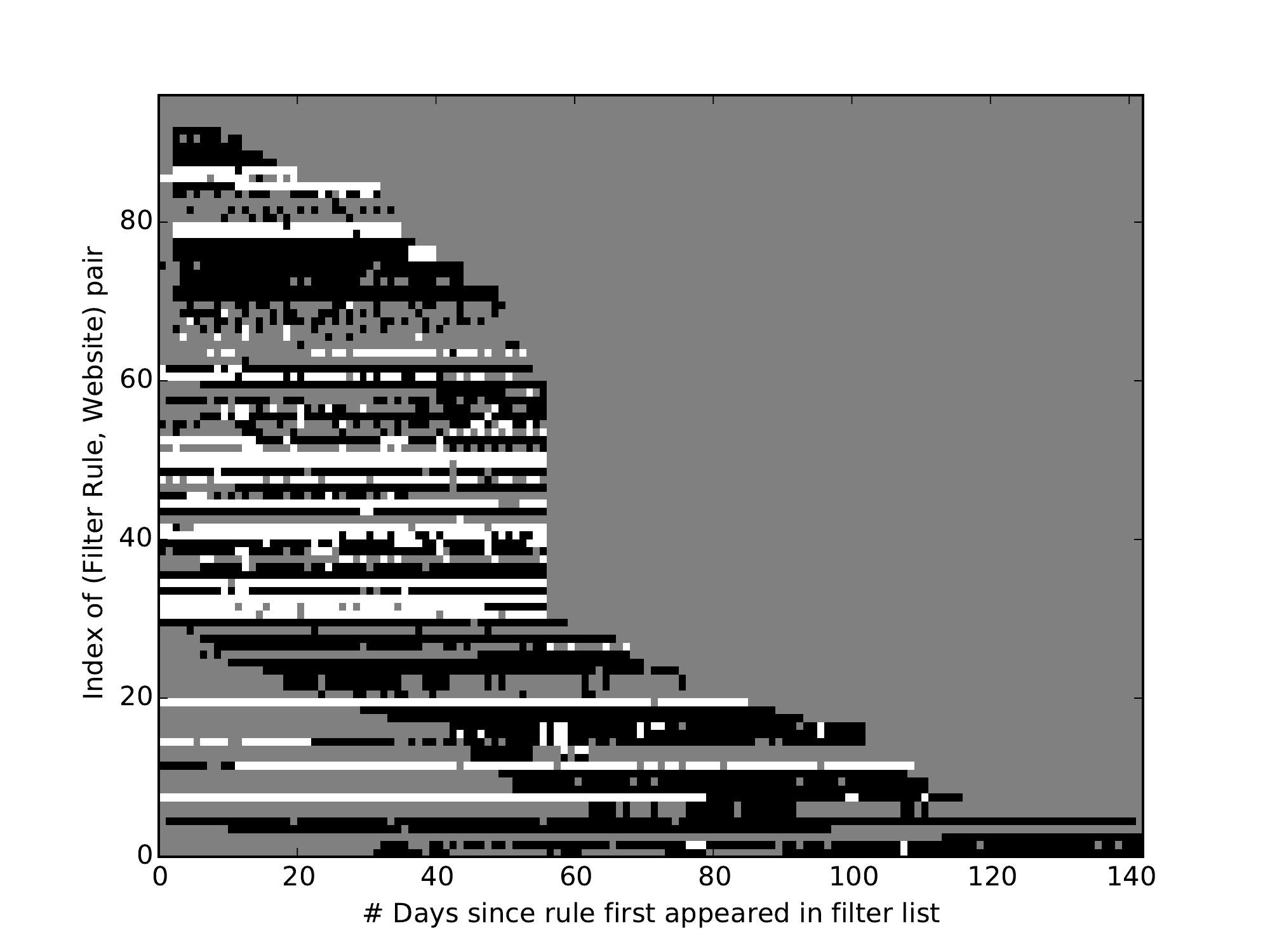}
\caption{Nano Defender effectiveness on websites targeted by Nano Defender}
\label{fig:combinedAbpUbpVsUbp}
\end{figure}

\subsection{Nano Defender}
We next examine the effectiveness of the more targeted anti-adblock
filter rules in the Nano Defender filter list. In addition to the Alexa
Top 5K (see Figure~\ref{fig:combinedAbpUbpVsAlexa:raw}) we also looked at
websites outside this list that were specifically targeted by Nano
Defender (see Figure~\ref{fig:combinedAbpUbpVsUbp}).

Figure~\ref{fig:combinedAbpUbpVsAlexa:avg} shows that
initially the filter effectiveness rate is approximated by a horizontal
line. The upward trend in the averages is overwhelmed by the growing confidence
intervals. Our best interpretation is that there is no
evidence of a decay in Nano Defender's filter rule effectiveness over time.

From roughly $x=45$ on, we no longer have sufficient data. This is
clearly illustrated in the increasing size of the confidence intervals,
and also apparent in
Figure~\ref{fig:combinedAbpUbpVsAlexa:raw}.
Notwithstanding this lack of data for later days, the absence of a
downward trend in the first part of
Figure~\ref{fig:combinedAbpUbpVsAlexa:avg} is significant, and thus we
conclude that in general, websites are currently neither tracking nor
responding to updates in the Nano Defender filter rules.

A curious possibility is that the absence of observed filter effectiveness decay in the Nano Defender data, versus the presence of decay in the EasyList data, controls for the possibility that web sites are simply drifting over time in their engineering practices. If that were the cause of our observed filter effectiveness decay, then we should see a similar effect in both data sets. Contrarily, the engineering of Nano Defender is much more specifically targeted than the engineering of EasyList, so it's also possible that a ``drift effect'' would impact EasyList's effectiveness more than it impacts Nano Defender.

\subsection{Comparing results with prior studies}

Iqbal et al.~\cite{ISQ17} also looked at subsections of EasyList and
Anti-Adblock Killer (their work predates the Nano Defender list). They
found, for the Alexa Top 5K websites, that Nano triggered on 8.7\%
websites and that AdblockWarningRemovalList combined with the
anti-adblock sections of EasyList (hereafter ``AWRL/EasyList'' ) only
triggered on 0.4\% websites. Nithyanand et al.~\cite{NKJ16} similarly
found anti-adblocking logic on 6.7\% of the Alexa Top 5K.

To compare their results with our data, where we have multiple samples
of each filter lists and of each website, we restate their question as
follows: Do {\em any} versions of Nano Defender list or AWRL/EasyList
trigger on {\em any} versions of each given website?
We find that the combined
Nano Defender list triggered on roughly 13.3\% of the Alexa Top 5K
(exactly 666 unique websites) and that AWRL/EasyList triggered on 0.06\%
of the Alexa Top 5K (exactly 3 websites). Our observed growth in
anti-adblocking logic relative to Iqbal's results likely combines two effects: a genuine increase in
websites using such logic, as well as increased engineering efforts on
the part of Nano Defender to detect and filter such logic. AWRL/EasyList
has remained comparatively static with fairly few commits in the same
period of time; websites and advertisers have clearly engineered around
AWRL/EasyList.




We next focus on PageFair, a commercial service that provides websites
with adblocking analytics and adblock-resistant advertising. Nithyanand
et al.~\cite{NKJ16} looked at a number of such services, finding 20 web
sites using PageFair, which was then successfully blocked by AdBlockPlus
and Privacy Badger, but not Ghostery. We detected 67 separate websites in
the Alexa Top 5K using PageFair, all of which are successfully filtered by the
Nano Defender list.

Nithyanand also discussed the arms race of
websites detecting and responding to adblockers. They noted that adblock
detection scripts are often loaded from popular content distribution networks
such as Cloudflare. One prominent such project is
``BlockAdblock''.
This
project appears to have been available from Cloudflare since at least
August 2015. We detected 20 websites in the Alexa Top 5K using some
variant of BlockAdblock. Surprisingly, we detected only one of
these websites using the suggested Cloudflare CDN URL; 8
of the remaining 19 used a URL owned by an advertising company, another 10 chose to serve a copy of the script
themselves, and one website used a different CDN service.
\section{Conclusions and future work}

We observed an approximate 0.13\% decrease per day in the effectiveness
on the Alexa Top 5000 web sites of new rules in the EasyList filter list in the
immediate period after they were added. However, websites did not
appear to be actively responding to updates in the more specialized
and less widely used Nano Defender list.

\proposeddelete{Taken together, these results strongly suggest that advertisers and
website owners are much less likely to care about these more
specialized, but less widely used, filter lists in comparison to the
widely used EasyList.}

There are numerous opportunities for additional work in this area.
\proposeddelete{
One such possibility would be to measure the decay in effectiveness of
filter list rules as they become more widely known and used. In
particular, if any other filter lists are ever added to the Adblock Plus
defaults, it would be an exceptional opportunity to compare how website
owners and advertisers react to filter lists suddenly spiking in usage.
Additionally, alternatives to AdblockPlus like uBlock
Origin\footnote{\url{https://github.com/gorhill/uBlock}} have begun
adding extensions to the EasyList
syntax\footnote{\url{https://github.com/gorhill/uBlock/wiki/Static-filter-syntax}}.
It might be worthwhile to attempt to replicate the findings in this
paper with these various alternatives and determine how the expanded
capabilities impact filter list decay over time.
Scaling up our methodology, e.g.~to the Alexa Top Million, would}
For example, scaling up our methodology to run on millions rather
than thousands of web sites, and for longer periods of time, would
certainly be feasible and interesting. Using Scrapinghub, scaling the
data collection is straightforward, albeit more expensive. 
Analyzing larger volumes of collected data would certainly require larger
computing clusters, which require an additional expense to rent. 
At least the process is straightforward to distribute on a cluster, since
each web site scrape, evaluated against each filter ruleset, is a completely independent task.

Another interesting possibility would be to cluster the various anti-adblocking
mechanisms that a longer-term survey might discover over time, to understand the
diversity of the anti-adblocking ecosystem. 



{\footnotesize \bibliographystyle{acm}
\bibliography{references}}

\end{document}